\documentclass[aps,prl,nobibnotes,nofootinbib,showpacs,reprint]{revtex4-1}

\usepackage{graphicx}
\usepackage{amsmath} 	

\newcommand{\vect}[1]{\mathbf{#1}}
\newcommand{\Dt}{\Delta t}
\newcommand{\Dx}{\Delta x} 
\newcommand{\svect}[1]{\mbox{\boldmath $#1$\unboldmath}}
\newcommand{\pol}{\svect{\varepsilon}}
\newcommand{\polp}{\svect{\varepsilon}_{p}}
\newcommand{\polx}{\svect{\varepsilon}_{x}}
\newcommand{\khx}{\vect{k}_{p}\cdot\hat{\vect{x}}}
\newcommand{\wpt}{ \omega_{p} t}
\newcommand{\p}{\hat{\vect{p}}_{x}}
\newcommand{\x}{\hat{\vect{x}}}
\newcommand{\pcl}{\vect{p}_{x}}
\newcommand{\xcl}{\vect{x}}

\newcommand{\wo}{\omega_{0}}
\newcommand{\Hnm}[2]{H^{'}_{#1#2}e^{ i\omega_{#1#2}t} }
\newcommand{\disp}[1]{$\displaystyle{#1}$}
\newcommand{\expsum}[2]{\sum_{#1=1}^{20}\frac{1}{#1!}{\left(#2\right)}^{#1}}
\newcommand{\expsumb}[2]{\sum_{#1=1}^{20}\frac{1}{#1!}{\left[#2\right]}^{#1}}
\newcommand{\ket}[1]{|#1\rangle}
\newcommand{\bra}[1]{\langle#1|}
\newcommand{\Fig}[1]{FIG.~\ref{#1}}
\newcommand{\Eq}[1]{Eq.~(\ref{#1})} 
\newcommand{\pulse}[1]{#1_{p}}
\newcommand{\vac}[1]{#1_{zpf}}
\newcommand{\mode}{{\vect{k},\lambda}}
\newcommand{\eps}{\epsilon_0}
\newcommand{\kdx}{\vect{k}\cdot\vect{x}}
\newcommand{\submode}{_{\vect{k},\lambda}}
\newcommand{\subrphase}{\tilde{\theta}_{\submode}}
\newcommand{\wt}{\omega t}
\newcommand{\kpdx}{\vect{k}_{p}\cdot\vect{x}}
\newcommand{\submodeup}{_{\vect{k},1}}
\newcommand{\submodedown}{_{\vect{k},2}}
\newcommand{\dkapa}{\Delta\kappa}

\begin{document}


\title{Quantized Excitation Spectrum of the Classical Harmonic Oscillator in Zero-Point Radiation}

\author{Wayne~Cheng-Wei~Huang}
\email{email: u910134@alumni.nthu.edu.tw} 

\author{Herman~Batelaan}
\email{email: hbatelaan2@unl.edu}

\affiliation{Department of Physics and Astronomy, University of Nebraska-Lincoln, Lincoln, Nebraska 68588, USA }

\begin{abstract}
	We report that upon excitation by a single pulse, the classical harmonic oscillator immersed in classical electromagnetic zero-point radiation, as described by random electrodynamics, exhibits a quantized excitation spectrum in agreement to that of the quantum harmonic oscillator. This numerical result is interesting in view of the generally accepted idea that classical theories do not support quantized energy spectra. 
\end{abstract}


\maketitle 


\begin{figure*}[t]
\centering
\scalebox{0.6}{\includegraphics{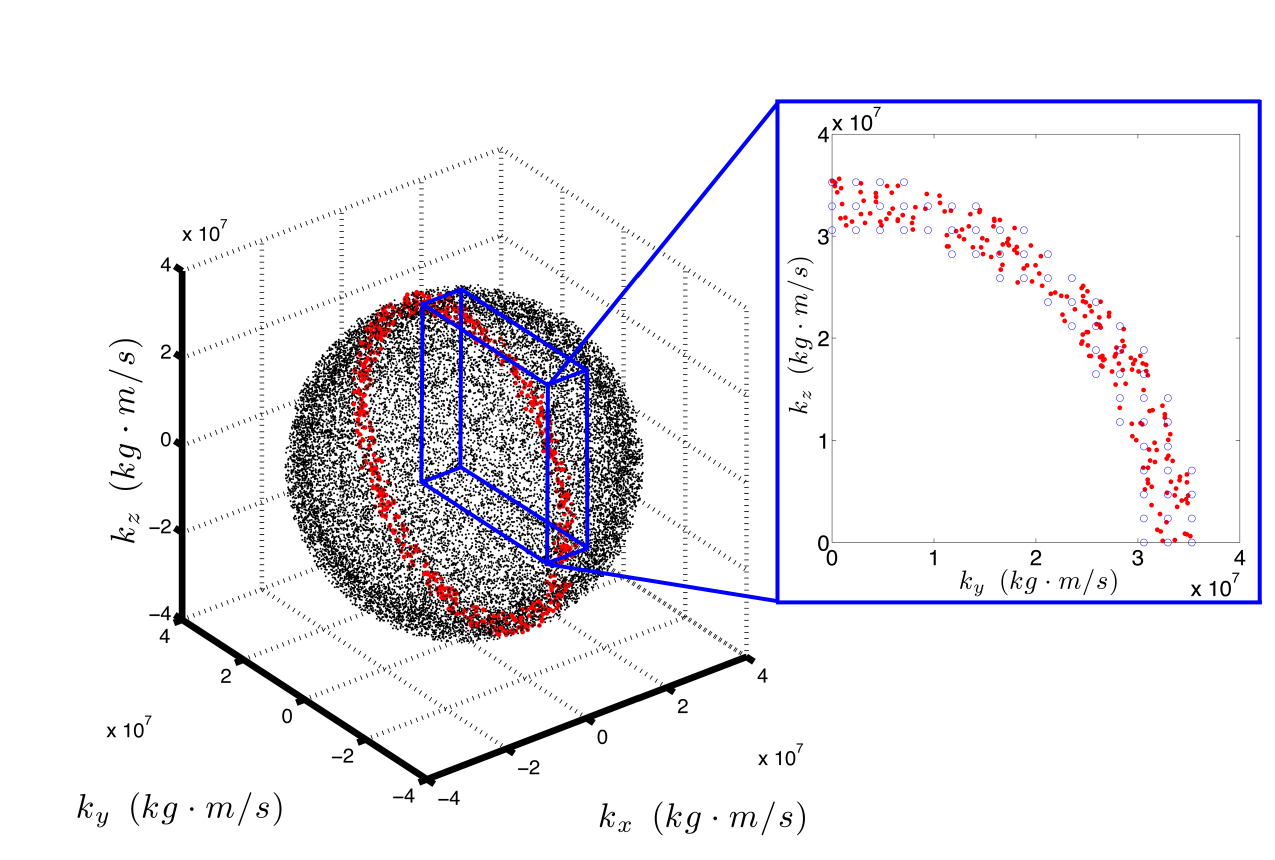}}
\caption{The sampled vacuum modes in $\vect{k}$-space. Left: The sampled vacuum modes (black dots) are distributed in $\vect{k}$-space as a spherical shell with thickness $\Delta/c$. The number of sampled modes shown is $N_{\omega} = 2 \times 10^4$. A slice of the spherical shell at $\vect{k}_{x} \simeq 0$ is highlighted (red dots). Right: A quarter of the highlighted slice (red dots) is projected on the $k_yk_z$-plane. The modes sampled in cartesian coordinates (red circles) are shown for comparison.}
\label{fig:vacmode_sampling} 
\end{figure*}

\begin{figure*}[t]
\centering
\scalebox{0.7}{\includegraphics{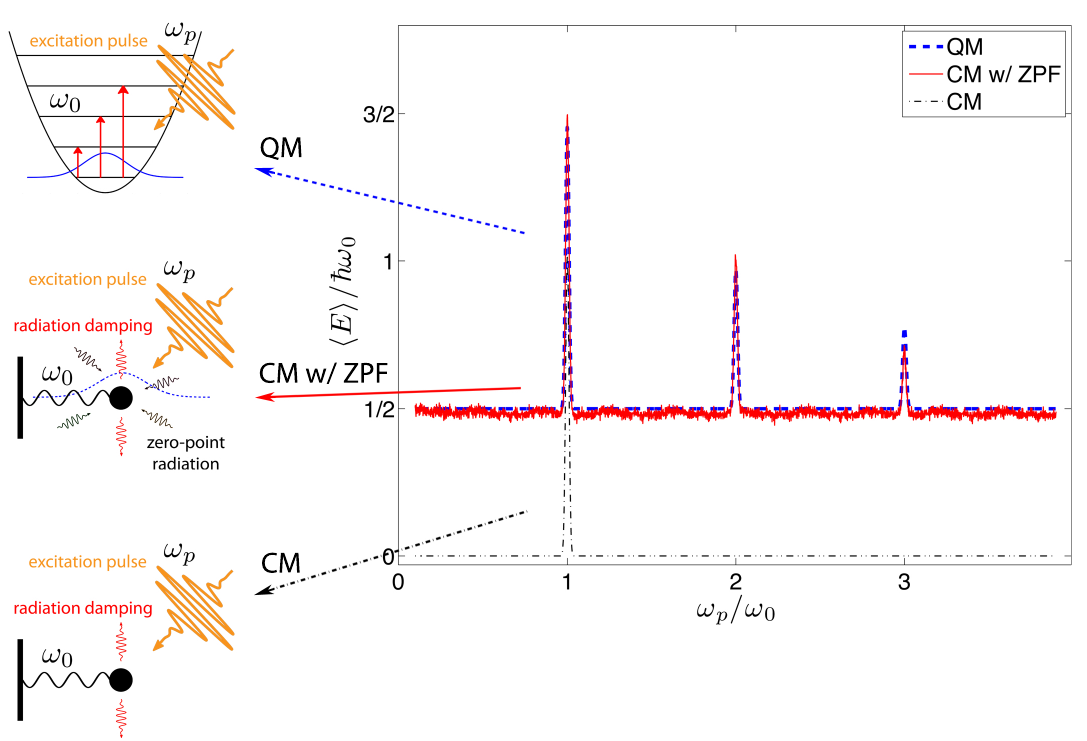}}
\caption{The excitation spectra of harmonic oscillators in different theories. The energy absorption spectrum is shown for the harmonic oscillator described in quantum mechanics (left top), classical ZPF theory (left middle), and classical mechanics (left bottom). The results from quantum mechanics and classical ZPF theory (i.e\ random electrodynamics) are in agreement. The vertical axis shows the ensemble averaged energy for the two classical theories and the expectation value of the energy for quantum mechanics. The energy is scaled by $\hbar\wo$. The horizontal axis shows the pulse frequency $\pulse{\omega}$ scaled by the oscillator's natural frequency $\wo$. The natural frequency is chosen to be $\wo = 10^{16}$(rad/s), the charge is chosen to be $q = 1.60 \times 10^{-19}$(C), and the mass is chosen to be $m = 9.11\times 10^{-35}$(kg). The choice of the mass is made to keep the integration time manageable without losing the physical characteristics of the problem. The ZPF frequency bandwidth is set as $\Delta = 2.2 \times 10^2\ \Gamma\wo^2$, which is much larger than the resonance width of a classical oscillator (i.e.\ $\Gamma\wo^2$). The calculations are convergent for increasing ZPF frequency bandwidth. The number of sampled ZPF modes is $N_{\omega} = 500$. Although this number of modes is much smaller than the one shown in \Fig{fig:vacmode_sampling}, $N_{\omega} = 500$ is sufficient for reaching numerical convergence.}
\label{fig:spectrum} 
\end{figure*}

\begin{figure*}[t]
\centering
\scalebox{0.5}{\includegraphics{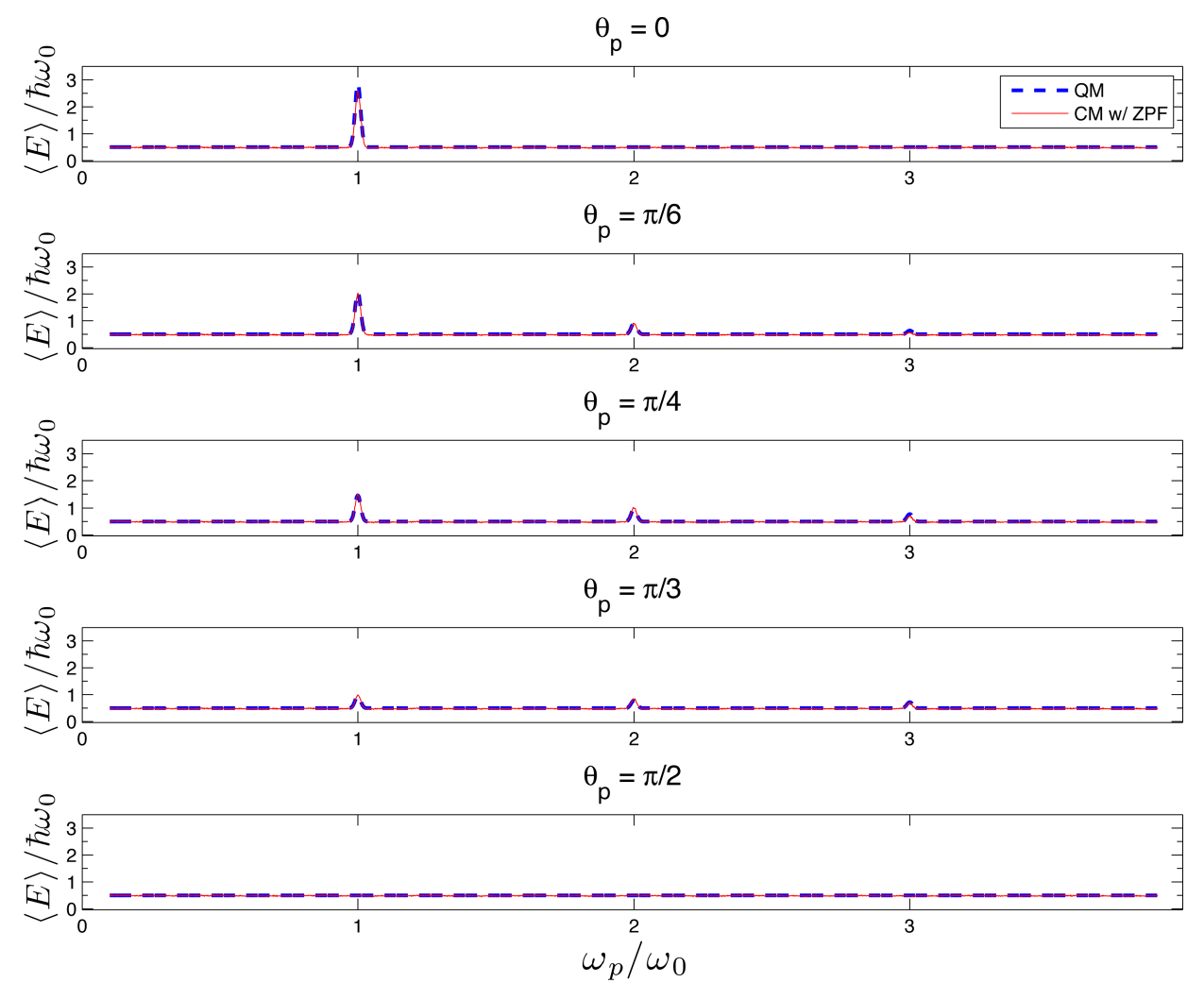}}
\caption{Excitation spectra of a harmonic oscillator at different excitation pulse angles $\pulse{\theta}$. The excitation spectra of a harmonic oscillator in classical ZPF theory (red solid line) are compared to those in quantum mechanics (blue dash line). For different excitation pulse angles $\pulse{\theta} = 0,\pi/6,\pi/4,\pi/3,\pi/2$, the spectra is shown. When the pulse is perpendicular to the direction of the oscillator's motion (i.e.\ $\pulse{\theta} = 0$), there is no higher harmonic resonance. When the pulse is parallel to the direction of the oscillator's motion (i.e.\ $\pulse{\theta} = \pi/2$), there is no excitation at all.}
\label{fig:pulse_angle} 
\end{figure*}

Historically, the discreteness of atomic spectra motivated the early development of quantum mechanics. Because quantum mechanics has made numerous successful predictions on atomic spectra, conventionally the discrete atomic spectrum is attributed to transitions between the quantized energy levels of an atom. Classical theories, such as stochastic electrodynamics (SED), are usually thought of as incapable of generating discrete atomic spectra, or discrete spectra for any potential that supports quantum mechanical bound states. Indeed, Peter Milonni has commented in his well-known book \emph{The Quantum Vacuum} that ``Being a purely classical theory of radiation and matter, SED is unable \ldots\ldots\ to account for the discrete energy levels of the interacting atoms.'' \cite{Milonni}. 

In the case of the harmonic oscillator, quantum mechanics predicts a discrete multi-resonance spectrum at the harmonics of the oscillator's natural frequency. The classical harmonic oscillator only supports a single resonance at the oscillator's natural frequency. In this Letter, we investigate the excitation spectrum of a one-dimensional harmonic oscillator immersed in the classical zero-point field, or zero-point radiation. The classical zero-point field (ZPF) is defined through the theory of random electrodynamics\footnote{
Random electrodynamics is also known as stochastic electrodynamics.} \cite{Boyer:RED, Boyer:RED&QM} as a nonzero homogenous solution to Maxwell's equations. The Plank constant $\hbar$ is brought into the classical theory as an overall factor that sets the strength of the classical zero-point field. Our simulation shows that a classical harmonic oscillator in the classical ZPF exhibits the same discrete multi-resonance excitation spectrum as a quantum harmonic oscillator. 

The excitation spectra of the classical and the quantum system are obtained through numerical computation of the total energy of the oscillator after excitation by a single pulse. The quantum mechanical result is obtained by numerically solving Schr\"{o}dinger's equation with a high-order multipole expansion. 

First, we discuss the theory for a one-dimensional quantum harmonic oscillator. The Hamiltonian for an oscillator interacting with a pulse is
\begin{equation}
¥	\widehat{H}_{qm} = \frac{ \left(\p - q\pulse{\vect{A}}(\x,t) \right)^2}{2m} + q\pulse{\phi}(\x,t) + \frac{m\wo^2}{2}\x^2,
\end{equation}¥
where $\p = (\hat{p}_{x},0,0)$, $\x = (\hat{x},0,0)$, $q$ is the charge, $m$ is the mass, $\wo$ is the natural frequency, and $(\pulse{\phi}, \pulse{\vect{A}})$ is the potential of the pulse. Using the Coulomb gauge $\nabla\cdot\pulse{\vect{A}} = 0$, Maxwell's equations yield $\pulse{\phi} = 0$ in the absence of external charge. In addition, $\p\cdot\pulse{\vect{A}} = \pulse{\vect{A}}\cdot\p$. Therefore, the Hamiltonian becomes
\begin{equation}\label{Hqm}
¥	\widehat{H}_{qm} = \left( \frac{\p^2}{2m} + \frac{m\wo^2}{2}\x^2 \right) - \frac{q}{2m}\left( 2\pulse{\vect{A}}\cdot\p - q\pulse{\vect{A}}^2 \right).	
\end{equation}¥
The eigenvalues of the first part of the Hamiltonian,
\begin{equation}
¥	\widehat{H}_{0} = \frac{\p^2}{2m} + \frac{m\wo^2}{2}\x^2,
\end{equation}¥
give the discrete energy levels $E_{n} = \hbar\wo(n+1/2) \equiv \hbar\omega_{n}$ for the unperturbed quantum harmonic oscillator. The oscillator is initially in its ground state. The second part of the Hamiltonian, 
\begin{equation}
¥	\widehat{H}^{'} = -\frac{q}{2m}\left( 2\pulse{\vect{A}}\cdot\p - q\pulse{\vect{A}}^2 \right),
\end{equation}¥
can induce transitions to the higher excited states. The field $\pulse{\vect{A}}$ is described by a propagating Gaussian pulse
\begin{equation}\label{qmApulse}
¥	\pulse{\vect{A}} = A_{0} \cos{( \khx - \wpt )} \exp{\left[- \left( \frac{\khx}{|\pulse{\vect{k}}|\Dx} - \frac{t}{\Dt} \right)^2 \right]}\polp,
\end{equation}¥
where $\Dt$ is the temporal width of the pulse, $\Dx = c\Dt$ is the spatial width, $\pulse{\vect{k}} = \pulse{\omega}/c\left( \sin{\pulse{\theta}}, 0, \cos{\pulse{\theta}} \right)$ is the wave vector of the carrier wave, and $\polp = \left( \cos{\pulse{\theta}}, 0, -\sin{\pulse{\theta}} \right)$ is the polarization vector of the pulse. The field amplitude is $A_{0}$. After excitation, the state $\ket{\psi}$(t) becomes a superposition of the energy eigenstates $\ket{n}$, 
\begin{equation}
¥	\ket{\psi}(t) = \sum_{n=1}^{N} c_{n}\ket{n}e^{-i\omega_{n}t}.
\end{equation}¥

To obtain the coefficients, $c_{n}$, of the excited state $\ket{\psi}$(t), we solve the Schr\"{o}dinger equation,
\begin{equation}
¥	\frac{d}{dt}C(t) = -\frac{i}{\hbar}\widehat{H}^{'}C(t),
\end{equation}¥
where $C(t)$ is a $N \times 1$ matrix and the Hamiltonian $\widehat{H}^{'}$ is represented by a $N\times N$ matrix,
\begin{equation}
¥	C =
	\begin{pmatrix}
	c_{1}(t) \\ c_{2}(t) \\ \vdots \\ c_{_{N}}(t)
	\end{pmatrix}, 	
	\widehat{H}^{'} = 
	\begin{pmatrix}
			&\hdots		&		\\
	\vdots	&\Hnm{n}{m}	&\vdots 	\\
			&\hdots		&		\\
	\end{pmatrix}.
\end{equation}
For each matrix element in the matrix of $\widehat{H}^{'}$, $\omega_{nm} = \omega_{n} - \omega_{m}$ and $H^{'}_{nm} = \bra{n}\widehat{H}^{'}\ket{m}$. To approximate a system with an infinite number of energy levels $N$, we choose $N = 20$ for numerical convergence. Since resonances at the harmonics of the oscillator's natural frequency are affected by the spatial dependence of the field, the dipole approximation $\vect{k}\cdot\x \simeq 1$ is not sufficient for the study of the excitation spectrum, making a numerical approach to this problem convenient. In our simulation, we expand the spatial dependence of the field $\pulse{\vect{A}} = \left( \widetilde{\vect{A}}_{p} + \widetilde{\vect{A}}_{p}^{\dagger} \right)/2$ up to the 20th order. Thus, the spatial dependence of the complex field
\begin{equation}
	\begin{split}
	¥	\widetilde{\vect{A}}_{p}  &= A_{0} \exp{ \left[ i( \khx - \wpt) \right]} \exp{\left[- \left( \frac{\khx}{|\pulse{\vect{k}}|\Dx} - \frac{t}{\Dt} \right)^2 \right]}\polp 		\\
					     	    &= A_{0} \exp{(-i\wpt)} \exp{\left[ -( t / \Dt)^2 \right]}f_{1}(\x)f_{2}(\x)f_{3}(\x)\polp,			     
	\end{split}¥
\end{equation}¥
is approximated by
\begin{align}
	¥	f_{1}(\x) &= \exp{(i\khx)} 		\nonumber \\
			     &\simeq \expsum{n}{i\khx},		\\
		f_{2}(\x) &= \exp{ \left[-\left( \frac{\khx}{|\pulse{\vect{k}}|\Dx} \right)^2 \right] }		\nonumber \\
			     &\simeq \expsumb{m}{-\left( \frac{\khx}{|\pulse{\vect{k}}|\Dx} \right)^2}, 	\\
		f_{3}(\x) &= \exp{\left[ 2\left(\frac{\khx}{|\pulse{\vect{k}}|\Dx}\right) \left(\frac{t}{\Dt}\right) \right]}		\nonumber \\
			     &\simeq \expsumb{k}{2\left(\frac{\khx}{|\pulse{\vect{k}}|\Dx}\right) \left(\frac{t}{\Dt}\right)}.	
\end{align}¥
Lastly, the matrix element of the operators, $\x$ and $\p$, are specified by 
\begin{align}
¥	\bra{n}\x\ket{m} &= \polx \sqrt{\frac{\hbar}{2m\wo}} \left( \sqrt{n}\delta_{m,n-1} + \sqrt{n+1}\delta_{m,n+1} \right), \\
	\bra{n}\p\ket{m} &= \polx i\sqrt{\frac{\hbar m\wo}{2}} \left( \sqrt{n}\delta_{m,n-1} - \sqrt{n+1}\delta_{m,n+1} \right),
\end{align}¥
where $\polx = (1,0,0)$. 

Next, we discuss the theory for an one-dimensional classical harmonic oscillator in the classical ZPF \cite{Boyer:RED&QM, Huang}. The classical Hamiltonian corresponding to the quantum system, \Eq{Hqm}, is
\begin{equation}
	\begin{split}
	¥	H_{cl} &= \left( \frac{\pcl^2}{2m} + \frac{m\wo^2}{2}\xcl^2 \right)		\\
	 		     &\quad - \frac{q}{2m}\left[ 2\left( \pulse{\vect{A}} + \vac{\vect{A}} \right)\cdot\pcl - q\left( \pulse{\vect{A}} + \vac{\vect{A}} \right)^2 \right].
	\end{split}¥	
\end{equation}¥
The classical ZPF in the Hamiltonian, $H_{cl}$, is specified by
\begin{equation}\label{Azpf}
¥	\vac{\vect{A}} =  \sum_{\mode} \sqrt{\frac{\hbar}{\eps V\omega}} \cos{( \kdx - \wt + \subrphase )} \pol_{\submode},
\end{equation}¥
where $\omega = c|\vect{k}|$, $\subrphase$ is the random phase uniformly distributed in $[0,2\pi]$, and $V$ is the volume of a cavity. The two unit vectors, $\pol_{\submodeup}$ and $\pol_{\submodedown}$, describe a mutually orthogonal polarization basis in a plane perpendicular to the wave vector $\vect{k}$. 
The pulse field $\pulse{\vect{A}}$ is identical to that in \Eq{qmApulse} except for $x$ being a classical quantity rather than an operator,
\begin{equation}
¥	\pulse{\vect{A}} = A_{0} \cos{( \kpdx - \wpt )} \exp{\left[- \left( \frac{\kpdx}{|\pulse{\vect{k}}|\Dx} - \frac{t}{\Dt} \right)^2 \right]}\polp.
\end{equation}¥
From the Hamiltonian, $H_{cl}$, the classical equation of motion can be derived,  
\begin{equation}
	\begin{split}
	¥	m\ddot{x}  &= -m\wo^2x + m\Gamma\dddot{x}	 \\
				&\quad + q\left[ ( \pulse{E}^{(x)} + \vac{E}^{(x)} ) + \left( \vect{v}_{x} \times ( \pulse{\vect{B}} + \vac{\vect{B}} ) \right)^{(x)} \right],
	\end{split}¥
\end{equation}¥
where $m\Gamma\dddot{x}$ is the phenomenological radiation damping term added to the theory. The damping coefficient is \disp{\Gamma \equiv \frac{2q^2}{3mc^3}\frac{1}{4\pi\eps}} \cite{Griffith}. Under the Coulomb gauge, the electric field is given by \disp{\vect{E} = -\frac{\partial \vect{A}}{\partial t}} and the magnetic field by \disp{\vect{B} = \nabla \times \vect{A}}. The symbol $E^{(x)}$ denotes the $x$-component of the vector $\vect{E}$. Because $\vect{v}_{x} = (v_{x},0,0)$, the magnetic part of the Lorentz force is zero. To avoid numerical runaway solutions, we assume the point-particle description of the charged particle and make the approximation $m\Gamma\dddot{x} \simeq-m\Gamma\wo^2\dot{x}$ \cite{Landau&Lifshitz, Jackson}. Hence, the equation of motion becomes
\begin{equation}
¥	m\ddot{x}  \simeq -m\wo^2x - m\Gamma\wo^2\dot{x}	 + q\left( \pulse{E}^{(x)} + \vac{E}^{(x)} \right),
\end{equation}¥
which will be used in our numerical simulation. 

To carry out the simulation, a set of modes $(\vect{k}_{i},\lambda)$ is chosen to construct the classical ZPF in \Eq{Azpf} \cite{Huang}. The wave vectors $\vect{k}_{i}$ have frequencies within the finite range $\left[ \wo-\Delta/2,\wo+\Delta/2 \right]$, where $\Delta$ is the ZPF frequency bandwidth and is much larger than the oscillator's resonance width $\Gamma\wo^2$. The ZPF modes in $\vect{k}$-space are sampled in spherical coordinates. Such a sampling method is computationally more efficient in reaching numerical convergence and approaches sampling in cartesian coordinates as $N_{\omega} \rightarrow \infty$ (see \Fig{fig:vacmode_sampling}). The specific steps of the sampling method are given in the following. For $i = 1\ldots N_{\omega}$,
\begin{equation}
¥	\vect{k}_{i}
	=
	\begin{pmatrix}
	k^{(x)}_{i} \\ k^{(y)}_{i} \\ k^{(z)}_{i} 
	\end{pmatrix}
	=
		\begin{pmatrix}
	k_{i}\sin{\theta_i}\cos{\phi_i} \\
	k_{i}\sin{\theta_i}\sin{\phi_i} \\
	k_{i}\cos{\theta_i}
	\end{pmatrix},
\end{equation}
where 
\begin{equation}
	\left\{ 
	\begin{array}{l}
	k_{i} = (3\kappa_{i})^{1/3} \\
	\theta_{i} = \cos^{-1}{(\vartheta_{i})} \\
	\phi_{i} = \varphi_{i},
	\end{array}
	\right.
	\end{equation}
	and
	\begin{equation}\label{mslct:layer3}
	\left\{ 
	\begin{array}{l}
	\kappa_{i} = (\wo-\Delta/2)^3/3c^3 + (i-1)\dkapa \\
	\vartheta_{i} = R_{i}^{(1)} \\
	\varphi_{i} = R_{i}^{(2)}.
	\end{array}
	\right.
\end{equation}
The stepsize $\dkapa$ is specified by
\begin{equation}
¥	\dkapa = \frac{\left[ (\wo+\Delta/2)^3/3c^3 - (\wo-\Delta/2)^3/3c^3 \right]}{N_{\omega}-1}.
\end{equation}¥
The random number $R^{(1)}$ is uniformly distributed in $[-1,1]$, and $R^{(2)}$ is uniformly distributed in $[0,2\pi]$. Finally, the polarization vectors are evaluated according to 
\begin{equation}
\begin{split}
&\pol_{\submodeup} = 
\begin{pmatrix}
\varepsilon_{\submodeup}^{(x)} \\ \varepsilon_{\submodeup}^{(y)} \\ \varepsilon_{\submodeup}^{(z)}
\end{pmatrix}
= 
\begin{pmatrix}
\cos{\theta_i}\cos{\phi_i}\cos{\chi_i}-\sin{\phi_i}\sin{\chi_i} \\
\cos{\theta_i}\sin{\phi_i}\cos{\chi_i}+\cos{\phi_i}\sin{\chi_i} \\
-\sin{\theta_i}\cos{\chi_i}
\end{pmatrix}       
\\	
\\			      
&\pol_{\submodedown} =  
\begin{pmatrix}
\varepsilon_{\submodedown}^{(x)} \\ \varepsilon_{\submodedown}^{(y)} \\ \varepsilon_{\submodedown}^{(z)}
\end{pmatrix}
=
\begin{pmatrix}
-\cos{\theta_i}\cos{\phi_i}\sin{\chi_i}-\sin{\phi_i}\cos{\chi_i} \\
-\cos{\theta_i}\sin{\phi_i}\sin{\chi_i}+\cos{\phi_i}\cos{\chi_i} \\
\sin{\theta_i}\sin{\chi_i}
\end{pmatrix}
.
\end{split}
\end{equation}
For a large number of modes, $N_{\omega}$, the sampling in spherical coordinates and cartesian coordinates approach each other, and the volume factor V in \Eq{Azpf} can be estimated by
\begin{equation}
V \simeq (2\pi)^3\frac{N_{\omega}}{V_\vect{k}}.
\end{equation}
Here, the  $\vect{k}$-space volume $V_{\vect{k}}$ encloses the sampled ZPF modes $\vect{k}_{i}$ in a spherical shell,  
\begin{equation}
V_{\vect{k}} = \frac{4\pi}{3}\left(\frac{\wo+\Delta/2}{c}\right)^3 - \frac{4\pi}{3}\left(\frac{\wo-\Delta/2}{c}\right)^3.
\end{equation}

The result of our numerical simulation for the excitation spectrum is given in \Fig{fig:spectrum}. When the classical ZPF is absent, the classical harmonic oscillator supports only a singe resonance at its natural frequency. With the classical ZPF acting as a constant background perturbation, the classical harmonic oscillator exhibits a discrete multi-resonance excitation spectrum with the background energy shifted up to $\hbar\wo/2$ \cite{Boyer:RED&QM,Huang}. The shape and the magnitude of the resonance peaks are in agreement with the quantum mechanical result. Furthermore, as the pulse angle $\pulse{\theta}$ changes from $0$ to $\pi/2$, the energy of the harmonic oscillator in the classical ZPF scales in the same way as that of the quantum harmonic oscillator, as shown in \Fig{fig:pulse_angle}. 

In conclusion, we have shown that the classical harmonic oscillator in the classical ZPF displays the same discrete multi-resonance excitation spectrum as the quantum harmonic oscillator. The agreement between classical ZPF theory and quantum mechanics is within the error of our numerical convergence. Extension of numerical work to atomic system appears to be interesting \cite{Cole:hydrogen}. Whereas in Cole's work the nonlinearity in the atomic $1/r$ potential alone affords the possibility of subharmonic excitation \cite{Cole:subharmonic}, in our work the nonlinearity is due to the existence of the classical ZPF and the spatial dependence of the excitation pulse, as exemplified by the necessity of calculation beyond the dipole approximation. Our numerical approach may also be useful to critically test recent claims that superposition and entanglement are supported by classical ZPF theories \cite{Cavalleri, de la Pena}. 

We gratefully acknowledge comments from Prof. Peter W.~Miloni. This work was completed utilizing the Holland Computing Center of the University of Nebraska and the Extreme Science and Engineering Discovery Environment (XSEDE), which is supported by NSF Grant No.~OCI-1053575. The funding support comes from NSF Grant No.~0969506.

\end{document}